\documentstyle[12pt]{article}

\input epsf

\newcommand{\ppp}[1]{%
        \setbox0=\hbox{#1}%
        \kern-.02em\copy0\kern-\wd0
        \kern+.04em\copy0\kern-\wd0
        \kern-.02em\raise.0217em\box0}
\newcommand{\vek}[1]{
         \mathchoice{\mbox{\boldmath$#1$}}%
        {\mbox{\boldmath$#1$}}%
        {\ppp{$\scriptstyle#1$}}%
        {\ppp{$\scriptscriptstyle#1$}}}

\newcommand{\lsim}{$\raisebox{-0.8ex} {$\stackrel{\textstyle <}{\sim}$}$}
\newcommand{\gsim}{$\raisebox{-0.8ex} {$\stackrel{\textstyle >}{\sim}$}$}

\begin{document}  
\begin{titlepage}
\vspace*{-2cm}
\begin{flushright}
\bf 
TUM/T39-98-1
\end{flushright}

\bigskip 

\begin{center}
{\large\bf Coherent Vector Meson Production from  Deuterons}

\vspace{2.cm}

{\large L.~Frankfurt$^{a,d}$, G.~Piller$^{b}$, M.~Sargsian$^{b,e}$, 
M.~Strikman$^{c,d}$}
\date{\today{}}

\vspace{2.cm}

${(a)}$ School of Physics and Astronomy, Tel Aviv University,
Tel Aviv,  69978 Israel
\\
${(b)}$ Physik Department, Technische Universit\"{a}t M\"{u}nchen,
D-85747 Garching, Germany 
\\
${(c)}$ Pennsylvania State University, University Park,
     PA 16802, USA
\\
$(d)$ Institute for Nuclear Physics, St. Petersburg, Russia
\\
${(e)}$ Yerevan Physics Institute, Yerevan 375036, Armenia

\vspace*{3cm}

{\bf Abstract}
\bigskip

\begin{minipage}{15cm}
In this paper we discuss 
coherent photo- and leptoproduction of vector mesons from 
deuterons at intermediate (virtual) photon energies, 
$ 3 \leq \nu \leq 30\,GeV$. We consider the 
scattering from unpolarized and polarized targets    
as well as processes where the polarization of the recoil deuteron 
is measured. Our main motivation results from the need 
for a quantitative analysis of the space-time structure 
of photon-induced processes. 
In this respect we suggest several possibilities to explore the 
characteristic longitudinal interaction length in coherent vector 
meson production. 
Furthermore, we outline methods for an investigation 
of color coherence effects. 
Besides the presentation of benchmark values for the maximal possible 
color coherence  effect  in various kinematic regions, 
we illustrate the anticipated phenomena within the color 
diffusion model. 
Finally we recall that the determination of vector meson-nucleon 
cross sections is not a closed issue. 
Besides being not known to  a satisfying accuracy, 
they can be used to determine 
the strength of the $D$-state in low mass vector mesons.

\end{minipage}

\end{center} 



\vspace*{1.cm}


\end{titlepage}

\section{Introduction}

Exclusive vector meson production from nucleons and nuclei 
can be used to investigate the transition from 
non-perturbative to perturbative strong interaction mechanisms.
To identify different kinematic domains we  consider the scattering 
of a (virtual) photon with four-momentum $q^{\mu}=(\nu,\vek q)$ 
from a nucleon. 
($Q^2 = -q^2$, $M$ stands for the invariant mass of the nucleon target,
and $x=Q^2/2M\nu$ is the Bjorken scaling variable in the lab frame.)
At $\nu \geq 3\,GeV$ and $x \leq 0.1$ one can decompose 
the amplitude for the production of a vector meson $V$  
in terms of hadronic Fock states of the 
(virtual) photon (see e.g. \cite{Gribov70,Spital76}):
\begin{equation}
f^{\gamma^*N\rightarrow VN} = 
\sum_{h}{<0|\epsilon_{\gamma^*}\cdot J^{em}|h> d\tau_{h}
\over E_{h}-\nu}f^{hN\rightarrow V N}.
\label{eq:spectral}
\end{equation}
Here $\epsilon_{\gamma^*}$ is the polarization vector of the 
virtual photon, and $J^{em}$ denotes the electromagnetic current. 
$E_{h}$ stands for  the energy of the intermediate hadronic state $h$ with 
phase space  $d\tau_{h}$.

At intermediate photon energies $3 \leq \nu \leq 30 \,GeV$ 
and low momentum transfers 
\linebreak
$Q^2 \,\lsim \,1~GeV^2$ 
contributions to $f^{\gamma^*N\rightarrow VN}$ from 
hadronic Fock states with large invariant mass $m_h$   
are  suppressed by large energy denominators 
$E_{h} - \nu  \approx  ({m_{h}^2 + Q^2})/({2 \nu})$ 
(and small photon-hadron couplings).
The restriction to light intermediate hadronic states  
leads to vector meson dominance (VMD).  
Here, as seen from the laboratory frame, the vector 
meson is formed  prior to the interaction with  the target 
(see e.g. \cite{Gribov70,BSYP}). 

This is different at large momentum transfers  
$Q^2\gg 1~GeV^2$ and $x \ll 0.1 $. 
Here it is legitimate to use closure over  intermediate 
hadronic  states $h$  and substitute         
the sum over hadronic states in Eq.(\ref{eq:spectral}) by  
quark-gluon Fock states which can be calculated 
in the framework of perturbative QCD. 
For example,  in the case  of longitudinally polarized photons  
short distance dominance  leads   at large $Q^2$
to the production of color-dipole  quark-antiquark pairs 
with small transverse size \cite{BFGMS,FKS,AFS,FGKSS}. 
Long after this initial hard interaction the 
final state vector meson is formed.

Interesting details about the production process  
can be obtained by embedding it into nuclei.   
Here details about the initially produced quark-gluon 
wave packet and the formation of the finally measured vector meson 
are probed via the interaction with spectator nucleons 
(for recent reviews see e.g. \cite{CT}).
In this context we discuss coherent photo- and leptoproduction 
of vector mesons from deuterons at photon energies 
$3 \leq \nu \leq 30\,GeV$ and $0\leq Q^2 \leq 10\,GeV^2$. 
In this kinematic regime one is most sensitive to the transition from 
non-perturbative to perturbative production mechanisms. 
The corresponding amplitudes can be split into two pieces:
a single scattering term in which only one nucleon participates 
in the interaction, and a  double scattering contribution.   
Here the photon interacts 
with one of the nucleons inside the target and produces an intermediate 
hadronic state which subsequently re-scatters from the second nucleon 
before forming the final state vector meson. 

In the considered process the following scales seem to be most 
relevant: the average {\bf transverse size} of the wave packet 
which for the case of longitudinal photons is
$b_{ej}\approx  4 \ldots 5/Q$ for the contribution 
of the minimal Fock space component at $Q^2 \,\gsim \,5 GeV^2$ \cite{FKS}.
For these $Q^2$ it amounts to less than 
a third of the typical diameter of a $\rho$ meson 
($b_{\rho}\approx 1.4\,fm$).
Furthermore, the initially produced small quark-gluon wave packet does 
not, in general, represent an eigenstate of the strong interaction 
Hamiltonian. The finally measured vector meson is formed after a typical 
{\bf formation time}  $\tau_f \approx 2\nu/\delta m_V^2$. 
Here $\delta m_V^2 \sim 1\,GeV^2$ is the characteristic 
difference of the squared  
masses of low-lying vector meson states. 
For energies $\nu \simeq 10\,GeV$ one finds $\tau_f \simeq 4\,fm$. 
Therefore in the considered kinematic region details about the expansion 
of the initially produced wave packet  are expected 
to significantly influence the production of vector mesons from 
nuclei. 
The kinematic dependence of the initial transverse 
size of the produced wave packet and the formation time 
of the final vector meson  suggest that 
contributions from re-scattering processes decrease  at large photon energies 
with rising $Q^2$.
This phenomenon is commonly called color coherence or color transparency
\cite{CT}.  

However dominant contributions  to high-energy, photon-induced 
processes result  from large longitudinal space-time intervals 
which increase with the photon energy \cite{Gribov70,GIP,LL}. 
In the considered kinematic domain 
the characteristic {\bf longitudinal interaction length}
$\lambda \approx 2\nu/(m_V^2 + Q^2)$
turns out to be of the order of typical nuclear dimensions 
and can have a major influence on the scattering process. 
A systematic investigation of the role of the longitudinal 
interaction length is most important, even for a qualitative 
understanding of photon-induced processes. 

In a previous publication \cite{FKMPSS97} we have focussed mainly 
on the derivation of vector meson production amplitudes. 
In this paper we collect several possibilities 
which can be used for a detailed 
investigation of the space-time structure of vector meson production.
Furthermore, we illustrate the anticipated color coherence effect 
in the framework of the quantum diffusion model \cite{FARRAR}. 
Finally we recall that the determination of vector meson-nucleon 
cross sections is not a closed issue. 
Despite many experimental studies they are not known to a satisfying 
accuracy. 
Furthermore, we point out that for polarized vector mesons  
they can be used to determine 
the strength of the $D$-state component in low mass vector mesons.

This paper is organized as follows: 
we recall in Sec.2 the photon-deuteron 
scattering amplitude. In Sec.3  we discuss various 
possibilities for an investigation 
of the longitudinal interaction length in 
coherent vector meson production.
Different signatures for color coherence are explored in Sec.4. 
In Sec.5 we comment on measurements of vector meson-nucleon 
cross sections. Finally we summarize.

\section{Photon-deuteron scattering amplitude}
\label{sec:amplitude} 

Let us briefly recall the vector meson production amplitude
for a deuteron target. 
Throughout this work we stay in  the laboratory frame 
where the deuteron with invariant mass $M_d$ is at rest.
Furthermore, we choose  the photon momentum 
in the longitudinal direction, i.e. 
$q^{\mu} = (\nu,{\bf 0}_{\perp},\sqrt{Q^2 + \nu^2})$. 
The single scattering, or Born amplitude reads \cite{FKMPSS97}:
\begin{eqnarray} \label{eq:ia}
F_{1}^j &=&  
f^{\gamma^*p\rightarrow Vp}(\vek l)\,
S_{d}^j\left(-{\vek l_{\perp}\over 2}, {l_{\_} \over 2}\right)~+~
f^{\gamma^*n\rightarrow Vn}(\vek l)\,
S_{d}^j\left({\vek l_{\perp}\over 2}, -{l_{\_} \over 2}\right). 
\end{eqnarray}
It is determined by the vector meson 
production amplitude $f^{\gamma^* p(n) \rightarrow V p(n)}$ from a  
proton or neutron, respectively, and the deuteron form factor $S_d^j$, 
where $j$ labels  the dependence on the target polarization. 
The transfered momentum  is denoted by 
$l^{\mu} = (l_0,\vek l) = q^{\mu} - k_V^{\mu}$ with $k_V$, the 
momentum of the produced vector meson.
Note that Eq.(\ref{eq:ia}) corresponds to the 
well known  result of Refs.\cite{FrGl66,bert} derived in the 
eikonal approximation, except for the presence of 
$l_{\_} = l_0 - l_z = \sqrt{M^2_{d}+\vek l^2}-M_d - l_z$   
in the form factor which accounts for the recoil of the 
two-nucleon system (for a derivation see \cite{FKMPSS97}). 

In the eikonal approximation the dominant contribution to the 
double scattering amplitude is given by \cite{FKMPSS97}:
\begin{eqnarray}   
F_{2}^{j} &\approx& 
{i\over 2} 
\sum_h \int {d^2 k_\perp  \over  (2\pi)^2}\,
S_{d}^j(\vek k_\perp,-\Delta_h) 
f^{\gamma^*N\rightarrow hN}
\left({\vek l_\perp\over 2}-\vek k_\perp\right) 
f^{hN\rightarrow VN}
\left({\vek l_\perp\over 2}+\vek k_\perp\right).
\label{eq:ds_lead}
\end{eqnarray}
Here $\Delta_h = {(Q^2 + 2m_h^2-m_{V}^2+t)/}{(4 \nu)}$, with 
the invariant mass $m_h$ of the intermediate hadronic state, 
and  $t = l^2$. 
In Eq.(\ref{eq:ds_lead}) we have neglected any isospin dependence of 
nucleon amplitudes.
 
The differential cross section 
for coherent vector meson production in (virtual) photon-deuteron 
interactions finally reads:
\begin{equation} \label{dif_crs_pho2}
\frac{d\sigma_{\gamma^{(*)}N}^{j}} {dt} =   \frac{1}{16\pi} 
\left( |F_{1}^j|^2 + 2 Re \left(F_{1}^{j *} F_{2}^j \right) 
+ |F_{2}^j|^2\right).
\end{equation}

We have not yet specified the polarization 
of the incident photon. 
Contrary to real photoproduction, where only transversely 
polarized photons ($\gamma_T$) enter, 
in leptoproduction also contributions from the exchange of 
longitudinal photons ($\gamma_L$) arise.
The leptoproduction cross section can therefore be split 
according to:
\begin{equation} \label{eq:prod_l}
\frac{d\sigma^j_{lN}}{dQ^2 d\nu dt} = \Gamma_V 
\frac{d\sigma^j_{\gamma^* N}}{dt} 
= \Gamma_V
\left(
\frac{d\sigma^j_{\gamma^*_T N }}{dt} 
+ \varepsilon \,\frac{d\sigma^j_{\gamma^*_L N}}{dt}\right). 
\end{equation}
Here $\Gamma_V$ is related to the flux of the virtual photon:
\begin{equation}
\Gamma_V = {\alpha_{em}\over 2\pi}{K\over Q^2}{1\over E_e^2}{1\over
  1-\epsilon},
\label{flux}
\end{equation}
where $\alpha_{em}=1/137$ is the electromagnetic coupling 
constant and $K=\nu(1-x)$. 
The parameter $\epsilon= (4 E_e E_e' - Q^2)/(2 (E_e^2 + E_e'^2) 
+ Q^2)$ specifies the photon polarization and is determined 
by the incoming and scattered lepton energies $E_e$ and 
$E_e'$. 

In the following we investigate the differential cross section 
${d\sigma_{\gamma^{(*)}N}^{j}}/{dt}$ for various target polarizations 
$j$. For this purpose we have to specify  the nucleon 
amplitudes  which enter in Eqs.(\ref{eq:ia},\ref{eq:ds_lead}).
In this respect we should mention that we assume 
$s$-channel helicity conservation throughout this work.

\subsection{Vector meson dominance} 
\label{sec:vmd}

In the kinematic domain of vector meson dominance 
($Q^2 \,\lsim \,1\,GeV^2$ and $\nu \geq 3\,GeV$)  
the vector meson production amplitude for transverse photons 
is successfully parametrized as \cite{BSYP}:
\begin{equation}
f^{\gamma^*_T N \rightarrow V N} 
\approx 
{\sqrt{\alpha_{em}\pi}\over g_{V}}{m^2_{V}\over m^2_{V}+Q^2}
f^{V_T N\rightarrow  V N}.
\label{eq:VMDT}
\end{equation} 
Here $V_{\mu}$ ($V=\rho,\omega,\phi$) stands for a  
vector meson  with invariant mass $m_V$ and coupling constant 
$g_V$, and $f^{V_T N\rightarrow  V N}$ denotes the amplitude 
for the elastic scattering of a transversely polarized vector meson 
from a  nucleon. 
The amplitude for longitudinal photons is given by:
\begin{equation}
f^{\gamma^*_L N\rightarrow  V N} = \xi {\sqrt{Q^2}\over m_V}
f^{\gamma^*_T N\rightarrow  V N}, 
\label{eq:VMDL}
\end{equation}
where $\xi =  f^{V_L N\rightarrow  VN}/f^{V_T N\rightarrow  V N}$ 
is the ratio of the longitudinal and transverse 
vector meson scattering amplitudes.

It remains to fix the amplitude $f^{\gamma^*_T N \rightarrow V N}$.
We illustrate most of the discussed issues for the case of 
$\rho$ production and use the parametrization:
\begin{equation} \label{eq:prod_ampl}
f^{\gamma^*_T  N \rightarrow \rho N} = 
A (i +  \alpha) e^{{B(Q^2)\over 2} t},  
\end{equation}
with 
$A = \sigma_{V_T N} \sqrt{\alpha_{em}\pi} m^2_{V}/((m^2_{V}+Q^2) g_{V})$. 
In real photoproduction of $\rho$ mesons one finds  
at $\nu = 17\,GeV$ 
typically $A \approx  68 \,\mu b$ \cite{BSYP}.
For the slope $B(Q^2)$ we use the empirical values 
\cite{BSYP,slopes}:
$B(Q^2 < 1\,GeV^2) = 7\,GeV^{-2}$, 
$B(1 < Q^2 < 2\,GeV^2) = 6\,GeV^{-2}$,  and  
$B(2 < Q^2 < 10\,GeV^2) = 5\,GeV^{-2}$. 
The $\rho$-nucleon amplitude is then obtained from Eq.(\ref{eq:VMDT}) 
using the  slope $B\approx 8~GeV^{-2}$. 
Furthermore we take $\xi^2 = 0.5$ \cite{BSYP}.
The real part of the amplitude in Eq.(\ref{eq:prod_ampl}) 
can be estimated 
at high energies through an empirical relation 
between the amplitudes for $\rho$ photoproduction and 
elastic pion-nucleon scattering:
\begin{equation}
f^{\gamma p \to \rho p} \sim f^{\pi^+ p \to \pi^+ p} 
+ f^{\pi^- p \to  \pi^- p}.
\label{pin}
\end{equation}
{}From a compilation of $\pi$-proton  scattering data 
at the considered energies one obtains $\alpha \approx -0.2$ 
\cite{piN}.
This corresponds to a decrease of the $\rho$-proton cross 
section with rising center of mass energies 
(see e.g. \cite{Crittenden}) as described by 
contributions from secondary Regge trajectories. 
Note that the situation is different for $\phi$ production. 
In the considered kinematic region the corresponding production 
cross section increases with the center of mass energy $s$.
{}From the Gribov relation \cite{Gribovrel},
\begin{equation}
Ref(s,t=0) = {\pi\over 2}{d \,Im f(s,t= 0)\over d \,\log s},
\label{realp}
\end{equation}   
one finds $\alpha =0.13$. 
In the following we neglect any $t$-dependence of $\alpha$.

It should be mentioned that 
Regge trajectories which correspond to isospin exchange    
are not important for isoscalar targets as the deuteron.
In the Born approximation such contributions vanish exactly.

\subsection{Quantum diffusion}

To illustrate the anticipated implications of color coherence 
in vector meson production at large $Q^2\gg 1\,GeV^2$ we employ 
the quantum diffusion model  \cite{FARRAR}.
It describes in a simplified manner the evolution and re-scattering 
of a quark-gluon wave packet which is produced in the initial 
(virtual) photon-nucleon interaction.
For $\rho$ production 
the corresponding re-scattering amplitude is given 
by \cite{FGMSS}:
\begin{equation} 
f^{ej}(z,t,Q^2) = i\sigma_{ej}(z,Q^2) 
\exp\left\{
{B\over 2}t + 
{b^2_{\rho}\over 6} 
\left({\sigma_{ej}(z,Q^2)\over \sigma_{\rho N}}-1\right) t
\right\}.
\label{cch1}
\end{equation}
The interaction cross section of the quark-gluon wave packet 
at a longitudinal distance $z$ from its initial production point is 
\cite{FARRAR}:
\begin{equation} 
\sigma_{ej}(z,Q^2) = \sigma_{\rho N}
\left\{\left(
{z\over \tau_f} +  
{b^2_{ej}(Q^2)\over  b^2_{\rho}}\left(1-{z\over \tau_f}\right)\
\right)
\Theta(\tau_f-z) + \Theta(z-\tau_f)\right\}.
\label{cch3}
\end{equation}
As already discussed in the introduction, $\sigma_{ej}$ 
is determined by the transverse size of the initially produced 
ejectile, $b_{ej}(Q^2)\sim 1/Q$, 
and the characteristic formation time $\tau_f$ 
of the vector meson. 
For distances larger than $\tau_f$ it coincides with the 
$\rho$ meson-nucleon cross section $\sigma_{\rho N} \approx 27\,mb$ 
\cite{BSYP}. 

Another class of models describes the production of vector 
mesons within a hadronic basis including  nondiagonal transitions 
between different intermediate hadronic states 
(see e.g. \cite{FRS,DS,KNNZ94}).
Here re-scattering is controlled by the 
destructive interference between diagonal (elastic) 
and nondiagonal (diffractive dissociation) transitions 
in meson-nucleon scattering processes. 
For the production of ground state vector mesons 
the results of such models are qualitatively 
similar to  the considered quantum diffusion picture.

\section{Probing the longitudinal interaction length}
\label{3a}

High energy  photon-induced processes are dominated  
by contributions from 
large longitudinal space-time intervals.
For coherent vector meson production characteristic 
distances can be extracted from the production  
amplitudes in Eqs.(\ref{eq:ia},\ref{eq:ds_lead})
after a Fourier transformation into coordinate space.   
The obtained length scales are to a good approximation  
inversely proportional 
to the longitudinal momentum transfers which enter in 
the form factors in Eqs.(\ref{eq:ia},\ref{eq:ds_lead}).  
For single scattering one finds:
\begin{equation} \label{eq:long_Born}
\delta_1 \sim \left|\frac{1}{l_{-}}\right| = 
\frac{2\nu}{Q^2 + m_V^2 - t}. 
\label{l1}
\end{equation}
Note that if the deuteron recoil is neglected  
one obtains $\delta_1\sim |1/l_z| = 
2 \nu /(Q^2 + m_V^2 - t - \nu t /2 M)$, which  
becomes constant at high energies. 
This, however, is 
in contradiction with our basic understanding of 
high-energy photon-induced processes being 
controlled by longitudinal distances which increase 
with the photon energy $\nu$ \cite{Gribov70,GIP,LL,Ioffe}. 

For double scattering we obtain from Eq.(\ref{eq:ds_lead}): 
\begin{equation} \label{eq:long_ds}
\delta_2 \sim \left|\frac{1}{-2\Delta_V}\right| 
= \left|\frac{2\nu}{Q^2 + 2 \langle m_h^2 \rangle - m_V^2 + t}\right|,  
\label{l2}
\end{equation}
where $\langle m_h^2 \rangle$ stands for the average squared mass  
of the intermediate hadronic states.

In the following we discuss several possibilities for 
a detailed investigation  of the characteristic longitudinal 
interaction length  in  vector meson production. 
For this purpose one has to concentrate on 
kinematic regions where distances of the order of the target size, 
$\delta_1, \delta_2 
\sim \langle r^2\rangle^{1/2}_d$, 
are reached.
Up to now investigations of this kind have not been carried 
out, although the presence of significant nuclear effects due to 
a non-vanishing longitudinal interaction length has been 
observed in shadowing effects in total photon-nucleus scattering 
cross sections (see e.g. \cite{Arneodo}). 
In particular we want to emphasize that the non-trivial 
$t$-dependence of the longitudinal interaction length 
in Eqs.(\ref{eq:long_Born},\ref{eq:long_ds}), 
which has been derived in Ref.\cite{FKMPSS97},  
needs to be verified to complete our understanding of photon-induced 
processes.

\subsection{Photoproduction from unpolarized deuterons}

At moderate photon energies $\nu \,\gsim \,3 \,GeV$, 
as available at TJNAF, the longitudinal interaction 
distance  for single scattering (\ref{eq:long_Born}) is for 
small values of the momentum transfer $t$  typically of 
the order of the deuteron size. 
For example at $\nu = 4\,GeV$ and $t = -0.2\,GeV^2$ 
one finds $\delta_1 \approx 2\,fm$.  
Consequently in this kinematic region, which is governed by the 
Born contribution,  
a strong dependence of the production cross section on 
$\nu$ is expected. 
To illustrate this issue consider the $\rho$ meson photoproduction cross 
section  (\ref{dif_crs_pho2}) calculated within vector meson dominance. 
It is important to use proper longitudinal interaction distances 
(\ref{eq:long_Born}, \ref{eq:long_ds}) as they appear 
in the scattering amplitude (\ref{eq:ia},\ref{eq:ds_lead}). 
We normalize our results by the production cross section  
obtained in the limit 
$\delta_1, \delta_2 \rightarrow \infty$. 
The corresponding ratio 
\begin{equation}
R_{\delta} = 
\left.
\frac{d\sigma_{\gamma N}}{dt}
\right/
\frac{d\sigma_{\gamma N}}{dt}(l_{-} = \Delta_{\rho} = 0).  
\label{Rl}
\end{equation}
is shown in Fig.1a for various photon energies. 
At $t \simeq -0.1 \,GeV^2$ we indeed observe 
a $15\%$ rise of $R_{\delta}$ for an increase of 
the photon energy from $3$ to $6\,GeV$.
At large $-t> 0.7\,GeV^2$, where double scattering dominates, only minor 
variations of the production cross section occur. 
This is due to the fact that the corresponding interaction length 
is larger than for single scattering.
Thus the difference in the  $t$-dependence of the longitudinal 
interaction length for single and double scattering 
reveals itself  through a different pattern in the energy 
dependence of the production cross section at small and large $|t|$.
For other low-mass vector mesons similar effects 
are expected.

\subsection {Polarized deuterons}

The coherent photoproduction of vector mesons from polarized 
deuterons or, equivalently, a measurement of the polarization 
of the recoil deuteron provides further possibilities to investigate 
effects due to a change of the characteristic longitudinal interaction 
length. If the polarization of the scattered or recoil deuteron, respectively, 
is chosen parallel to the photon momentum $\vek q$ one finds for 
the single scattering cross section:

\begin{eqnarray} \label{eq:rho_01}
\left.\frac{d\sigma^{\parallel}_{\gamma N}}{dt}
\right|_{Born} 
\sim
\left(F_C(\tilde l/2) - \frac{F_Q(\tilde l/2)}{\sqrt{2}}\right)^2 
+  
\frac{3 l^2_{-}}{2 \tilde l^2} F_Q(\tilde l/2) 
\left( 2\sqrt{2} F_C(\tilde l/2) + F_Q(\tilde l/2) \right),
\end{eqnarray}
where $\tilde l^2 = \vek l_{\perp}^2 + l_{-}^2$.  
$F_C$ and $F_Q$ are the deuteron monopole and quadrupole 
form factors \cite{BJ}. 
The first term in Eq.(\ref{eq:rho_01}) vanishes at 
$\tilde l \approx 0.7\,GeV$, while the second, which is 
proportional to $l_-^2\sim 1/\delta_{1}^2$, 
decreases for increasing photon energies. 
As a consequence the Born contribution to the production cross section 
develops a node at $t\approx -0.5\,GeV^2$. Therefore in this 
kinematic region a strong energy dependence of the production 
cross section is expected. 
Note that the energy dependence of the second term in 
Eq.(\ref{eq:rho_01}) is a consequence of the particular 
choice of the deuteron polarization being parallel to the 
photon momentum.

In Fig.2  we show the differential  $\rho$ meson photoproduction cross 
section calculated within vector meson dominance for 
various photon energies. 
The significant energy dependence of the Born 
contribution at $t\approx  - 0.5 \,GeV^2$ leads  
to a decrease of the full production cross section at 
$t \approx  -0.35\,GeV^2$ of more than two orders of magnitude,   
if the  photon energy rises from $\nu = 3$ to $20\,GeV$.
In previous  treatments of photoproduction processes 
at small $t$ the longitudinal interaction length was taken to be proportional 
to the inverse momentum transfer $\delta_1 \sim |1/l_z|$ 
neglecting any $t$-dependence \cite{BSYP}. 
Within this approximation the above discussed node in the differential 
cross section  would be absent, even at large photon energies. 
Therefore polarized deuteron targets or, equivalently, a 
measurement of the polarization of the recoil deuteron 
allow a detailed investigation of the kinematic dependence of the 
longitudinal interaction length.

\subsection{Leptoproduction from unpolarized deuterons}

In high-energy leptoproduction processes the longitudinal interaction 
lengths $\delta_{1,2}$ (\ref{eq:long_ds},\ref{eq:long_Born})  
exhibit  a characteristic $Q^2$-dependence. 
At moderate values of $0.5 \lsim \,Q^2 \,\lsim \,2 \,GeV^2$,  
as available at TJNAF, variations of $\delta_{1,2}$ 
of the order of the target size are expected.
Therefore a strong $Q^2$-dependence of 
vector meson production is anticipated.
In the framework of vector meson dominance 
this is visualized in Fig.1b. 
We show the cross section ratio $R_{\delta}$ from Eq.(\ref{Rl}) 
for various $Q^2$ at a fixed photon energy $\nu = 6\,GeV$.
Indeed we observe at $ t \approx - 0.1\,GeV^2$ 
a decrease of $R_{\delta}$ by approximately $50\%$, 
if $Q^2$ rises from $0.5$ to $2\,GeV^2$.

Similar effects are expected at higher photon energies. 
Then -- to keep the typical longitudinal interaction length 
close to the target size -- larger 
values of $Q^2$ are needed.  
For example at HERMES energies, $\nu \approx 25 \,GeV$, 
values of  $Q^2 \approx (2-10)\,GeV^2$ are required. 
To avoid complications due to color coherence effects one has to 
focus on the kinematic domain 
where the Born contribution dominates, i.e. $-t\,\lsim \,0.4\,GeV^2$.

\section{Search for color coherence}

In this section we focus on the possibility to 
study color coherence effects in coherent 
vector meson production from deuterons. 
For this purpose one needs to investigate the $Q^2$-dependence of the 
corresponding cross section in kinematic regions 
where it is most sensitive to re-scattering.
However it is mandatory to account for possible modifications 
due to  changes of the characteristic longitudinal interaction 
lengths  $\delta_1$ and  $\delta_2$ from 
Eqs.(\ref{l1},\ref{l2}). 
Or, even better, the kinematics should be chosen such 
that these length scales stay reasonably constant.
For $Q^2>1\,GeV^2 > m_V^2,|t|$ this can be achieved 
in an approximate way 
by keeping the Bjorken scaling variable fixed.
It should be mentioned that in vector meson production 
the onset of color coherence might 
occur already at moderate values of $Q^2 \,\gsim \,1\,GeV^2$. 
This expectation is based on the observation that the vector 
current correlation function assumes its perturbative 
form already at relatively large distances \cite{Shuryak}.
Furthermore, large size fluctuations of  
the minimal wave function of longitudinally 
polarized photons are suppressed by a factor $1/Q^2$ 
as compared to the transverse case (see e.g. \cite{BFGMS}). 
Therefore color coherence effects should arise earlier 
in the longitudinal channel.  
 
We illustrate our investigations for the case of $\rho$ meson production. 
The obtained results can be applied, however,  to other vector mesons too.  
Differences in  the onset of color coherence for various 
vector mesons and their excited states are expected 
(see e.g. \cite{FARRAR}). 
They should result from differences in the mass scales which are involved in 
the hard interaction, as well as from deviations between the wave functions 
of the final state vector mesons. 
An experimental investigation of these  issues is certainly needed.

\subsection{Unpolarized deuteron targets}

The maximal signal which can result from color coherence is determined  
by the difference between the Born cross section,  which results 
from single scattering only, and the full production 
cross section at $Q^2\approx 0$.
In Fig.3 we show the ratio of the full  $\rho$ production 
cross section and the  Born term for various values of $t$ 
at $x=0.1$:
\begin{equation} \label{eq:Rct}
R_{ct} = \left.\left(\frac{d\sigma_{\gamma^*N}}{dt}\right)\right/ 
\left(\frac{d\sigma_{\gamma^*N}}{dt}\right)_{Born}. 
\end{equation}
We have applied the quantum diffusion model from Sec.2.2 
using a characteristic formation time $\tau_f$ as determined 
by the mass difference $\delta m_{\rho}^2 \approx 0.7\,GeV^2$.   
Color coherence effects are calculated for two different scenarios:
(i) color coherence applies to transversely and longitudinally  
polarized photons equally; 
(ii) color coherence affects only the longitudinal 
cross section but not the transverse one. 
In the second case the cross section ratio is calculated for 
a photon polarizations (\ref{eq:prod_l}) 
characterized by  $\varepsilon = 0.5$.

At small values of $t  \simeq -0.1\,GeV^2$ single scattering 
dominates and color coherence effects are small. 
Double scattering starts to become relevant already 
at moderate $t\,\gsim \, -0.2 \,GeV^2$. 
Here the interference of the double and single scattering amplitude 
is important. 
For real photoproduction it results in a cross section 
smaller than the Born piece.
Consequently here color coherence  leads to an 
increase of the cross section ratio $R_{ct}$ with rising $Q^2$.
At large $|t|$ coherent photoproduction of  
vector mesons is dominated by 
double scattering.
Therefore in this region color coherence implies a decrease of $R_{ct}$ 
at large $Q^2$. 
Note that even if we omit  color coherence 
effects for transversely polarized photons, 
the $Q^2$-dependence of the virtual photoproduction cross section 
remains nearly unchanged.

The difference between the full cross section and the Born 
piece can be enhanced  if one considers the ratio of   
cross sections at large and moderate $|t|$. 
Here any $Q^2$-dependence, 
apart from being caused by color coherence, cancels to a large 
extent.  
In Fig.4 we present the ratio of the $\rho$ meson production 
cross sections taken  at $x=0.1$ for 
$t = -0.4\,GeV^2$ and $t=-0.8\,GeV^2$. 
The Born cross section and the full vector meson dominance 
calculation differ by a factor four,  
leaving reasonable room for an investigation of color coherence.
We also present results 
obtained from the quantum diffusion model, 
assuming similar color coherence effects for 
longitudinal and transverse photons. 
To emphasize the strong dependence of the measured cross section 
ratio on the expansion mechanism of the initial wave packet 
we vary $\delta m_{\rho}^2$ between $0.7$ and $1.1\,GeV^2$.   
The observed sensitivity is due to the moderate photon energies considered.
On the other hand  variations of the initial size of the 
wave packet do not change our results significantly, 
e.g. a $30\%$ rise in $b_{ej}$ changes the cross section ratio 
over the entire range of $Q^2$ by less than $5\%$.

\subsection{Polarized deuterons}

At moderate values of $|t|$ vector meson production from unpolarized 
deuterons is largely dominated by the Born contribution.
If the polarization of the target, or alternatively, of 
the scattered deuteron is fixed  even here  
kinematic windows exist where the Born contribution is small and 
the production cross section becomes sensitive to re-scattering.

A promising signature for color coherence can be obtained by 
choosing the polarization of the target or the scattered 
deuteron, respectively, perpendicular to the vector meson production plane, 
i.e. parallel to $\vek q\times \vek l$.
The corresponding Born cross section is then proportional to: 
\begin{equation}
\left.\frac{d \sigma^{\perp}_{\gamma^*N}}{dt}\right|_{Born} 
\sim 
\left(F_C^2(\tilde l/2 ) - \frac{1}{\sqrt{2}} F_Q(\tilde l/2) 
\right) ^2,
\label{deny}
\end{equation}
irrespective of the magnitude of the longitudinal momentum 
transfer $l_{-}$. 
Consequently a node in the 
Born contribution occurs for $\tilde l = 0.7\,GeV$,  
which corresponds at large photon energies approximately 
to  $t \approx - 0.5\,GeV^2$.
In Fig.5 we show the $t$-dependence of the $\rho$ production cross 
section normalized to its value at $t=t_{min}$ for 
various values of $Q^2$ but fixed $x=0.1$.
We present results from vector meson dominance and quantum diffusion.  
With rising $Q^2$ one observes a considerable decrease of the production 
cross section at $-t\,\gsim \,0.4$, and a shift of the cross section 
minimum towards the node in the single scattering cross section 
at $t\approx -0.5\,GeV^2$.

Finally we consider the cross section for tensor polarization 
normalized by the unpolarized production cross section:
\begin{equation}\label{eq:tensor_pol}
A_d = \frac{d\sigma^{m=1}/dt + d\sigma^{m=-1}/dt - 
      2 d\sigma^{m=0}/dt}
      {d\sigma/dt}.
\end{equation}
We choose the spin quantization axis either 
perpendicular to the vector meson production plane  
or parallel to the photon momentum. 
In comparison to the corresponding asymmetry for the Born contribution 
we find a significant sensitivity to double scattering 
at $-t \geq 0.6\,GeV^2$ as demonstrated in Fig.6.
Here color coherence leads to a sign-change of $A_d$.

\section{Vector meson-nucleon cross sections}

We have illustrated above how coherent vector meson 
production from deuterons can be used for a detailed  investigation  
of color coherence effects. 
In a similar way it can be employed at low $Q^2$ to determine   
the interaction cross sections of  vector mesons.  
Their precise knowledge is relevant for various topics in strong 
interaction physics, ranging from the investigation of 
hadron production processes to the analysis of nucleus-nucleus collisions.  
Despite a long history of experimental studies 
(see e.g. \cite{BSYP,Crittenden}) 
$\omega$-, $\phi$- and $\psi$-nucleon cross sections  are not 
known to satisfactory accuracy. 
In particular we should mention that in the case of $\psi$-mesons 
and probably to some extend also $\phi$ mesons 
their extraction from photoproduction processes off nucleons in 
the framework of vector meson dominance is likely to give 
too small values (for an early discussion see e.g. \cite{FS85}).

This situation can be improved through an investigation 
of coherent vector meson production at low $Q^2 \, \lsim \, 0.5\,GeV^2$ 
in kinematic regions dominated by double scattering processes 
which have been discussed in the previous section. 
While unpolarized targets require a large momentum transfer, polarized 
targets allow such investigations already at moderate $t$ 
(see Sec.4.2). 
For illustration we discuss $\phi$ meson production. 
In Fig.7  we show the polarized photoproduction  
cross section $\sigma^{\perp}_{\gamma N}/dt$ normalized to its value at 
$t = t_{min}$ for two different 
$\phi$-nucleon interaction cross sections 
$\sigma_{\phi N} = 10$ and $15 \, mb$. 
The photon energy is fixed at $6 \,GeV$. 
At $t \simeq - 0.45\,GeV^2$, where double scattering is important, 
we observe a significant sensitivity on $\sigma_{\phi N}$. 
Here both production cross sections differ approximately by
one order of magnitude.

Finally let us mention that vector meson-nucleon 
cross sections for polarized vector mesons carry 
interesting information too. 
In the framework of quark models mesons are considered 
as bound states of a quark and antiquark (see e.g. \cite{IK}).  
Since vector mesons carry spin $1$, 
the $q\bar q$ pair can be in a $S$- or  $D$-state.
The latter, however, is of  asymmetric shape. 
In leptoproduction  a smaller than average configuration of the 
vector meson wave function is selected due to the involved mass 
difference. 
If the probed wave function component contains a significant 
$D$-state contribution, a  difference in the 
re-scattering process for longitudinally and transversely 
polarized vector mesons is expected. 
Thus polarized leptoproduction of vector mesons at low 
$Q^2 \,\lsim\,0.5 GeV^2$  
can give interesting information about the strength of 
the corresponding $D$-state component.

\section{Summary}

The space-time structure of exclusive production processes 
contains valuable information on the transition from
non-perturbative to perturbative  
strong interaction mechanisms. 
Especially the use of nuclear targets is of advantage 
since here the production process is probed at  
a typical length scale of around $2 \,fm$,  
as given by the average nucleon-nucleon distance.  

We have outlined several possibilities to investigate 
the characteristic longitudinal interaction length 
of photon-induced processes. 
Such a study has never been performed on a quantitative 
level. 
In this context we have stressed in particular the non-trivial 
dependence of the longitudinal interaction length on the momentum 
transfer $t$. 
This genuine new effect could be explored to great detail 
at TJNAF -- especially with polarized targets or 
deuteron polarimeters (e.g. at TJNAF Hall C).

Furthermore, we have explored possibilities to investigate 
color coherence effects.
Promising signatures have been presented  at moderate and 
large momentum transfers. 
We have illustrated the anticipated color coherence effects within 
the color diffusion model.
At the considered  energies a large sensitivity to details 
on the formation of the final vector meson has been observed.  

Finally we recall that the determination of vector meson-nucleon 
cross sections is not a closed issue. 
Besides being not known to a satisfying accuracy, 
they can be used to determine 
the strength of the $D$-state in low mass vector meson states.

\bigskip
\section*{Acknowledgments}
We would like to thank G. van der Steenhoven and Eric Voutier 
for discussions. 

\noindent
This work was supported  in part by the German-Israeli
Foundation Grant GIF -I-299.095, the U.S. Department of Energy
under Contract No. DE-FG02-93ER40771, and   the BMBF.
M. Sargsian is grateful to the Alexander von Humboldt Foundation 
for support. 
Three of us (G.P., M.S. and M.S.) enjoyed the kind hospitality 
of the Institute of Nuclear Theory, Seattle, during the 
preparation of this paper.


\newpage
\noindent
{\bf\large{Figure Captions:}}

\bigskip
\bigskip
\noindent
Figure 1:\newline
The cross section ratio $R_{\delta}$ from Eq.(\ref{Rl}) 
for photoproduction (a) for various values of $\nu$. 
$R_{\delta}$ for leptoproduction (b) 
for various values of $Q^2$.

\bigskip
\bigskip
\noindent
Figure 2:\newline
The differential cross section 
$d\sigma_{\gamma N}^{\parallel}/dt$ for the photoproduction 
of $\rho$ mesons off deuterons polarized parallel to the 
photon momentum.
The solid curves show  results from vector meson 
dominance using proper longitudinal interaction lengths 
(\ref{eq:long_Born},\ref{eq:long_ds}). 
The dashed  lines correspond to a longitudinal interaction 
length $\delta_1 \sim |1/l_z|$ instead of $\delta_1 \sim |1/l_-|$.
In both cases the cross sections decrease for increasing 
photon energies.

\bigskip
\bigskip
\noindent
Figure 3:\newline
The $Q^2$-dependence of the cross section ratio $R_{ct}$ 
from Eq.(\ref{eq:Rct}) for various values of $t$. 
The full line corresponds to vector meson dominance. 
The dashed curve results from the application of color diffusion 
to longitudinally and transversely polarized photons. 
For the dot-dashed lines color diffusion has been 
applied to longitudinal photons only. 
In this case 
the photon polarization parameter has been fixed at 
$\varepsilon = 0.5$.

\bigskip
\bigskip
\noindent
Figure 4:\newline
The $Q^2$-dependence of the cross section ratio for 
coherent $\rho$ production from unpolarized 
deuterons taken at $t = -0.8$ and $-0.4 \,GeV^2$.  
The solid line represents the complete vector meson dominance 
calculation. The dashed curve accounts for the Born 
contribution only. 
The shaded area exhibits results from  quantum diffusion 
(\ref{cch1}). The characteristic expansion time 
$\tau_f = 2 \nu/\delta m_\rho^2$ has been varied 
via $\delta m_\rho^2 = (0.7-1.1)\,GeV^2$.

\bigskip
\bigskip
\noindent
Figure 5:\newline
The cross section $d\sigma_{\gamma^*N}^{\perp}/dt$ normalized 
to its value at $t=t_{min}$ for various values of $Q^2$ and  
fixed $x=0.1$.
The solid lines present the complete vector meson dominance 
calculation. 
Results from quantum diffusion are  shown by the dashed curves.
At $t\simeq - 0.4 \, GeV^2$ 
both cross sections decrease with rising values of $Q^2$.
The arrow indicates the position of the node in the 
Born contribution.

\bigskip
\bigskip
\noindent
Figure 6:\newline
The tensor polarization asymmetry $A_d$ from 
(\ref{eq:tensor_pol}) for various values of $Q^2$ and fixed 
$x=0.1$.  
The spin quantization axis has been chosen either 
perpendicular to the vector meson production plane (a), 
or parallel to the photon momentum (b). 
The solid lines show the complete vector meson dominance 
calculation. The dotted curve represent the Born contributions. 
Results from quantum diffusion are  shown by the dashed curves.
At $-t > 0.4\, GeV^2$ the curves to the right correspond 
to increasing values of $Q^2$.

\bigskip
\bigskip
\noindent
Figure 7:\newline
The $t$-dependence (a) of the $\phi$ photoproduction cross section 
$d\sigma_{\gamma N}^{\perp}/dt$ 
normalized to its value at $t=t_{min}$ for different values 
of the  $\phi$-nucleon cross section 
$\sigma_{\phi N} = 15$ (full) and  
$10 \,mb$ (dashed). 
The photon energy is fixed at $\nu = 6\,GeV$. 
In (b) the ratio of both cross sections is shown.


\newpage
\begin{figure}
\centerline{\epsfysize=7.0truein\epsffile{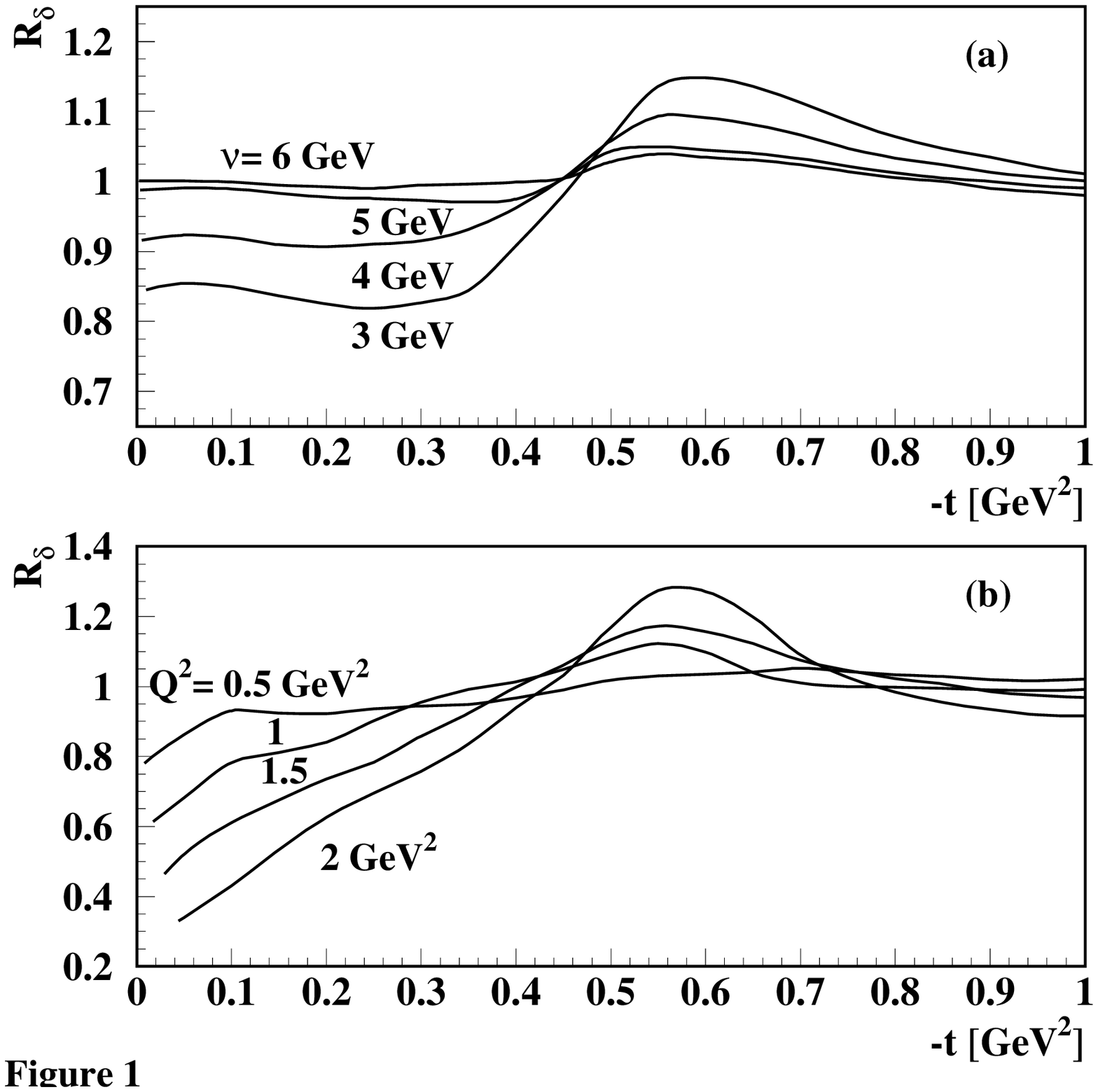}}
\label{Fig.1}
\end{figure}

\newpage
\begin{figure}
\centerline{\epsfysize=7.0truein\epsffile{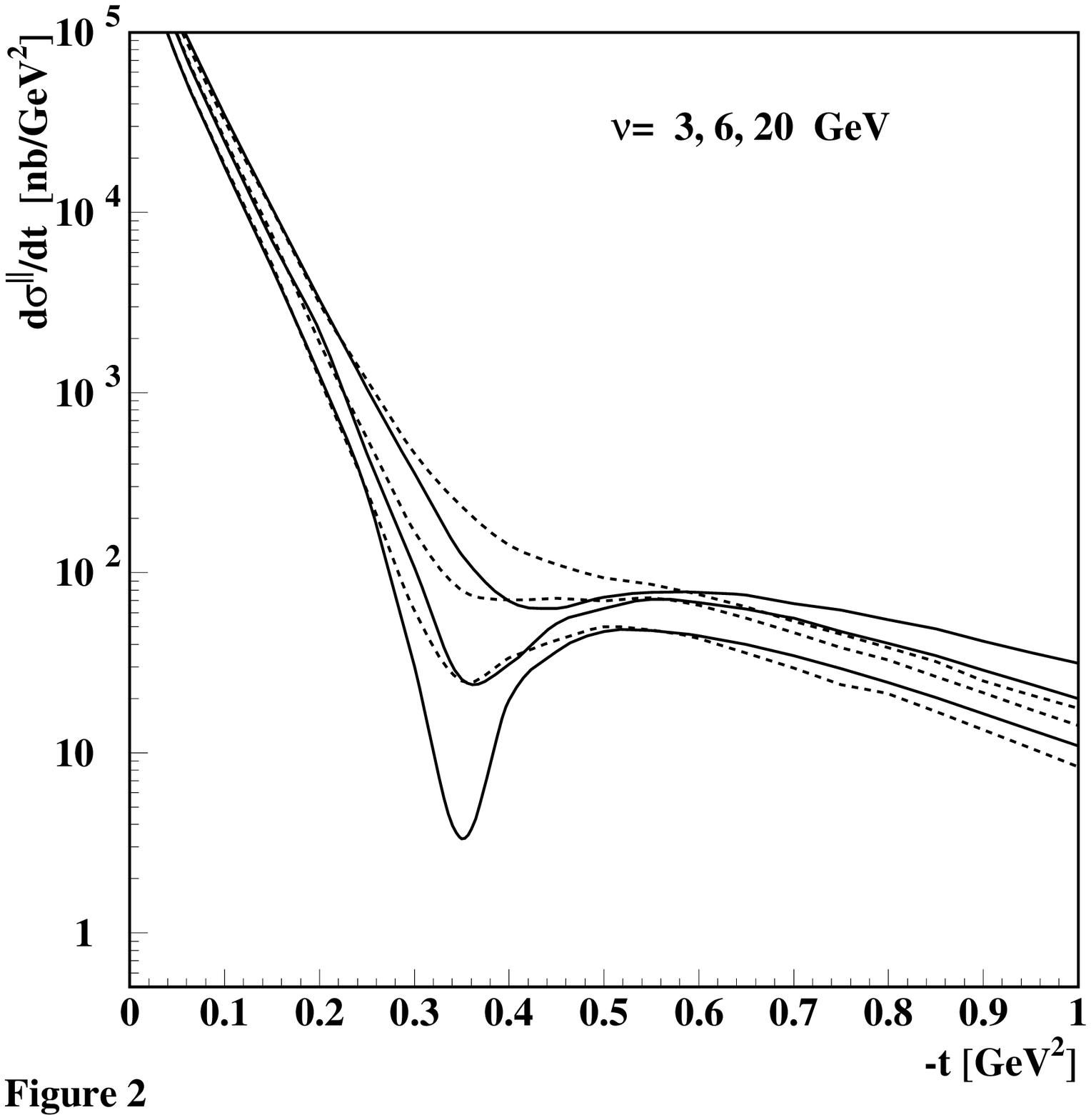}}
\label{Fig.2}
\end{figure}

\newpage
\begin{figure}
\centerline{\epsfysize=7.0truein\epsffile{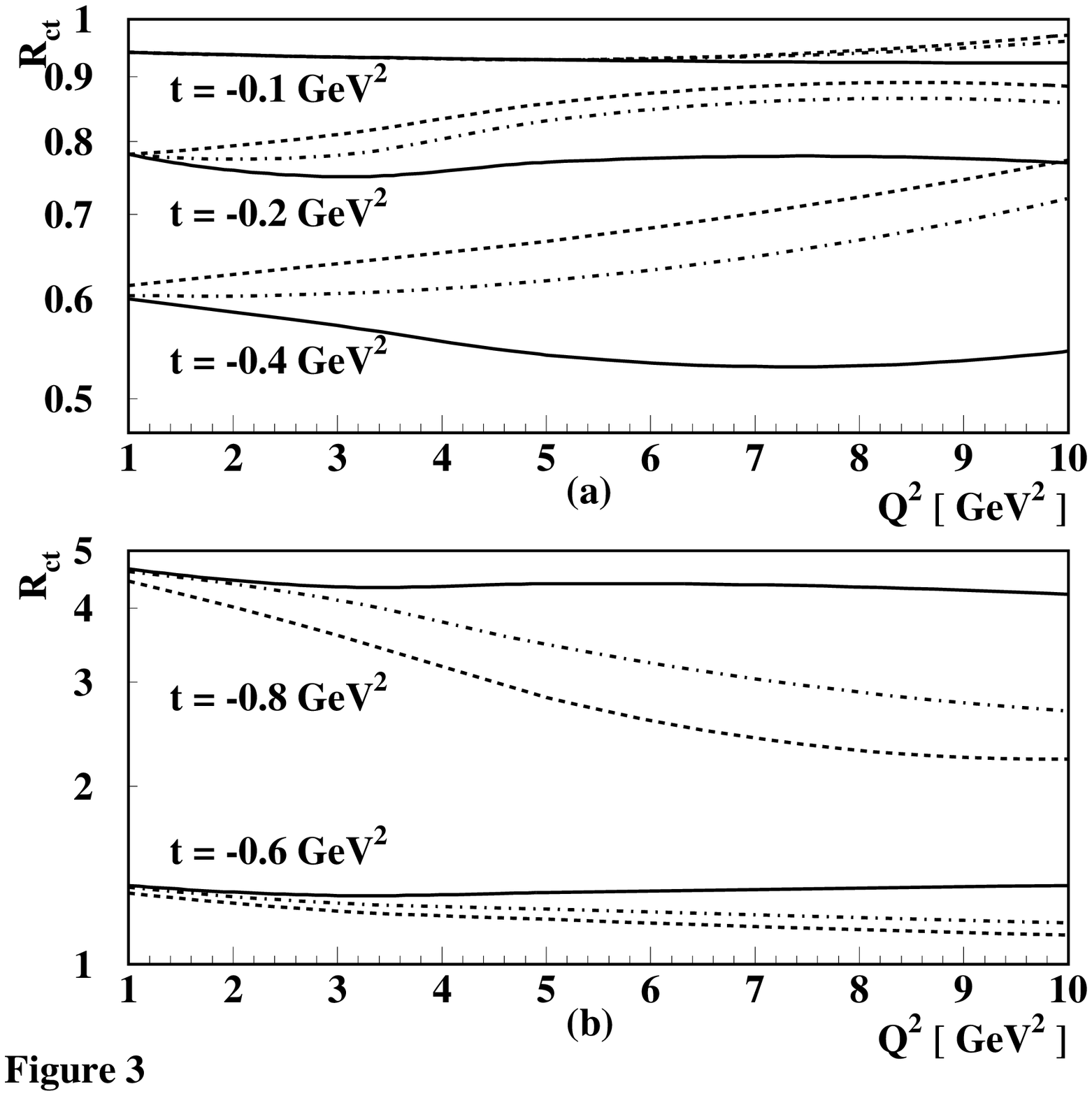}}
\label{Fig.3}
\end{figure}

\newpage
\begin{figure}
\centerline{\epsfysize=7.0truein\epsffile{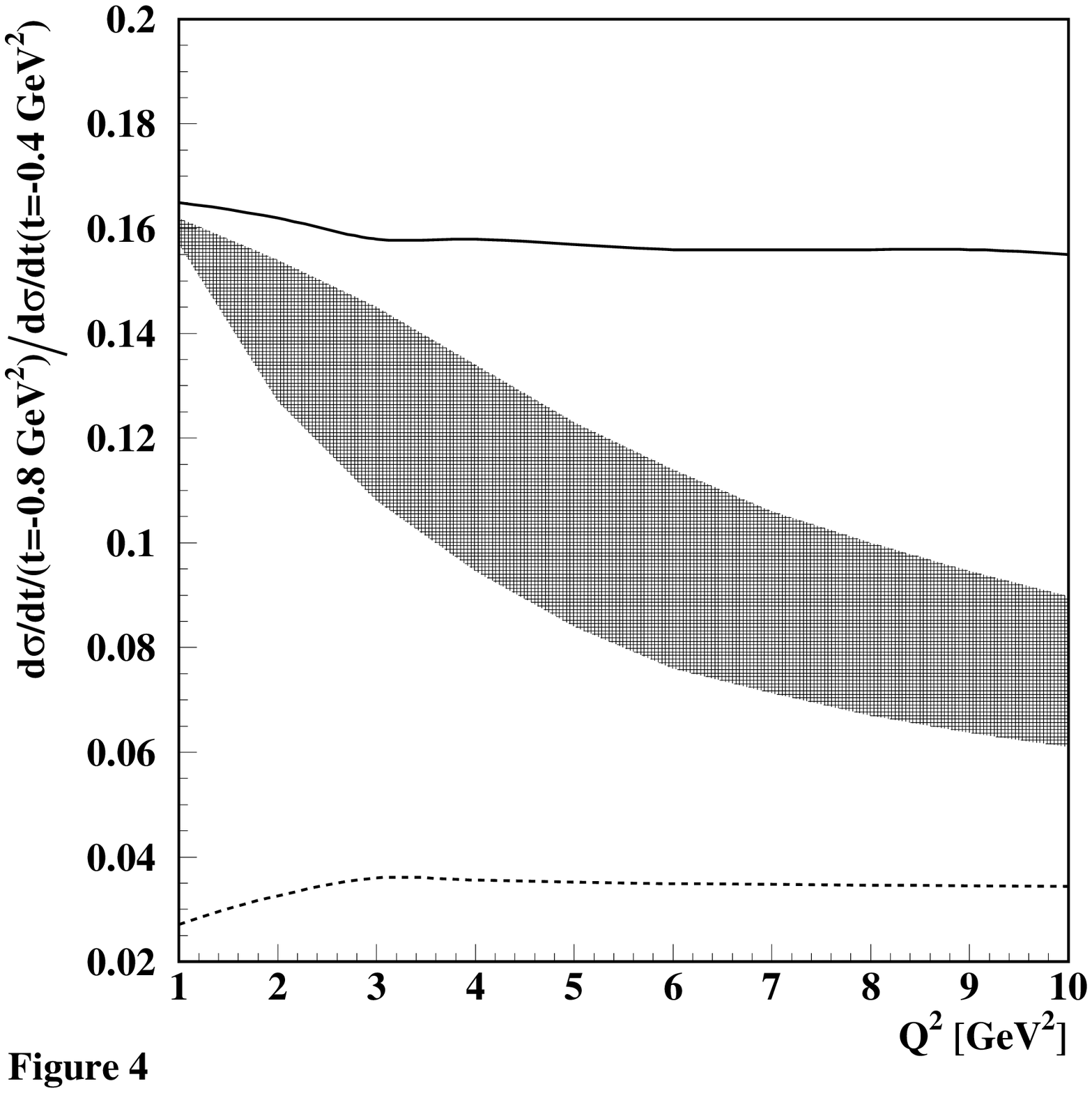}}
\label{Fig.4}
\end{figure}

\newpage
\begin{figure}
\centerline{\epsfysize=7.0truein\epsffile{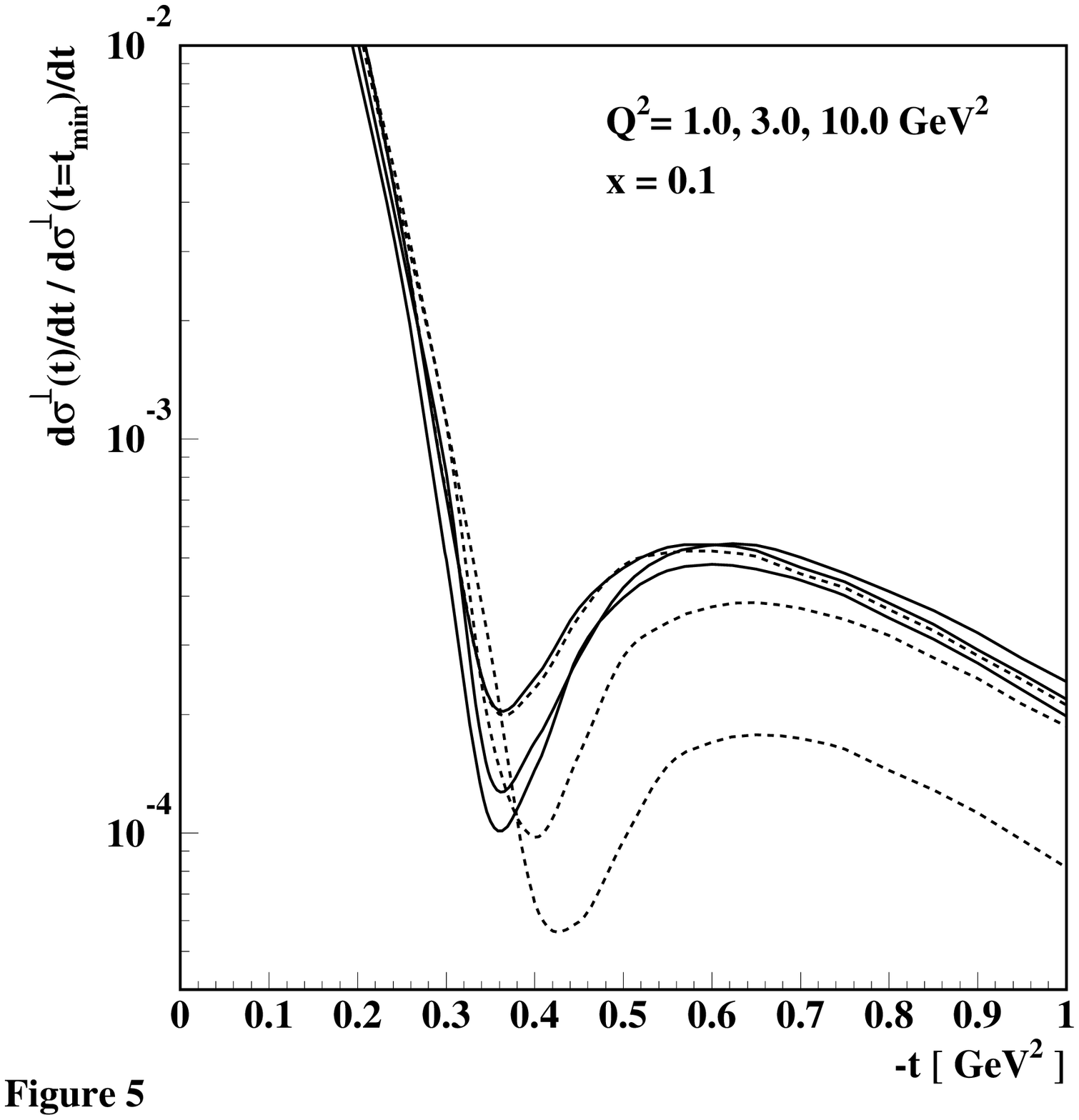}}
\label{Fig.5}
\end{figure}

\newpage
\begin{figure}
\centerline{\epsfysize=7.0truein\epsffile{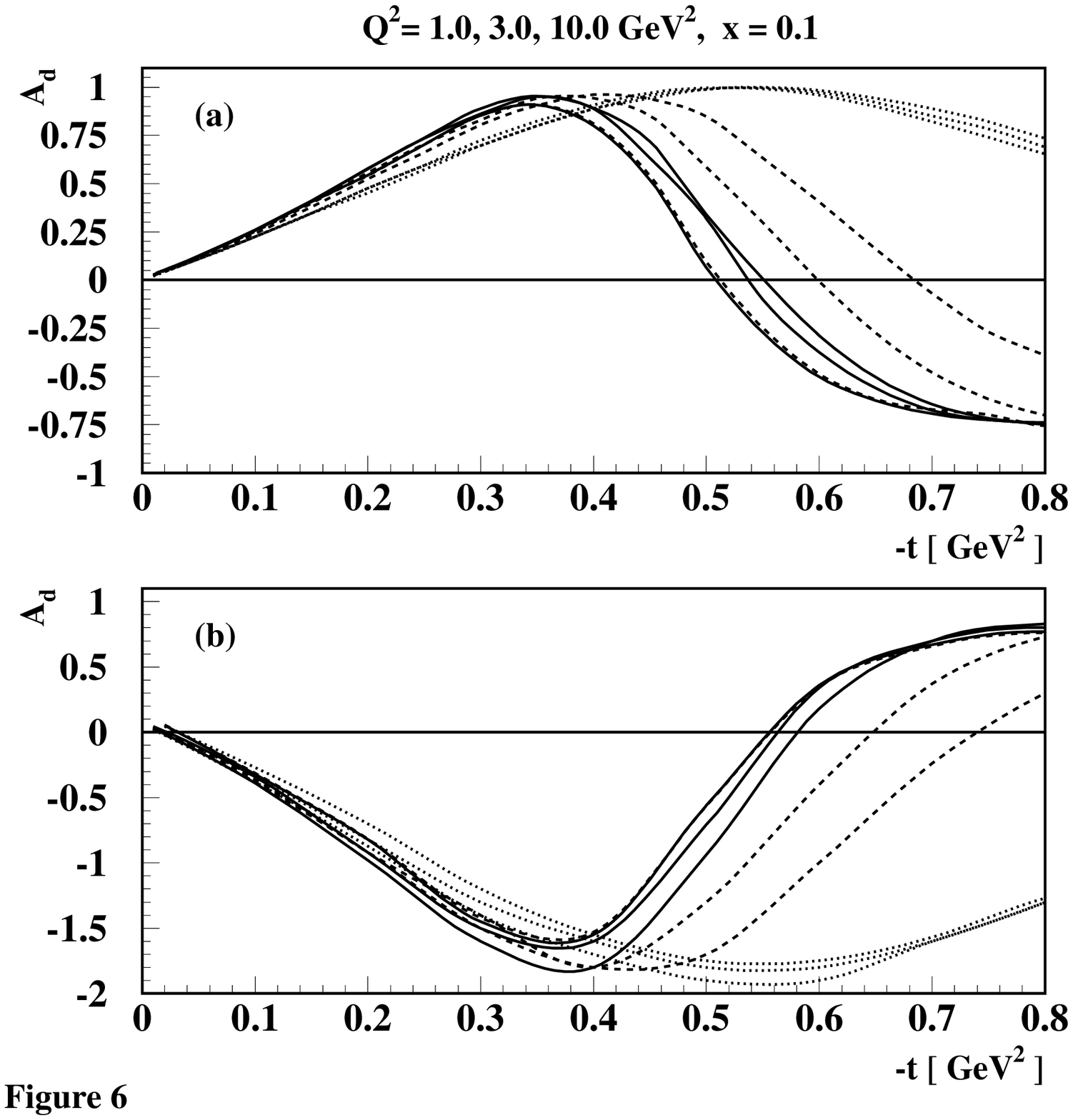}}
\label{Fig.6}
\end{figure}

\newpage
\begin{figure}
\centerline{\epsfysize=7.0truein\epsffile{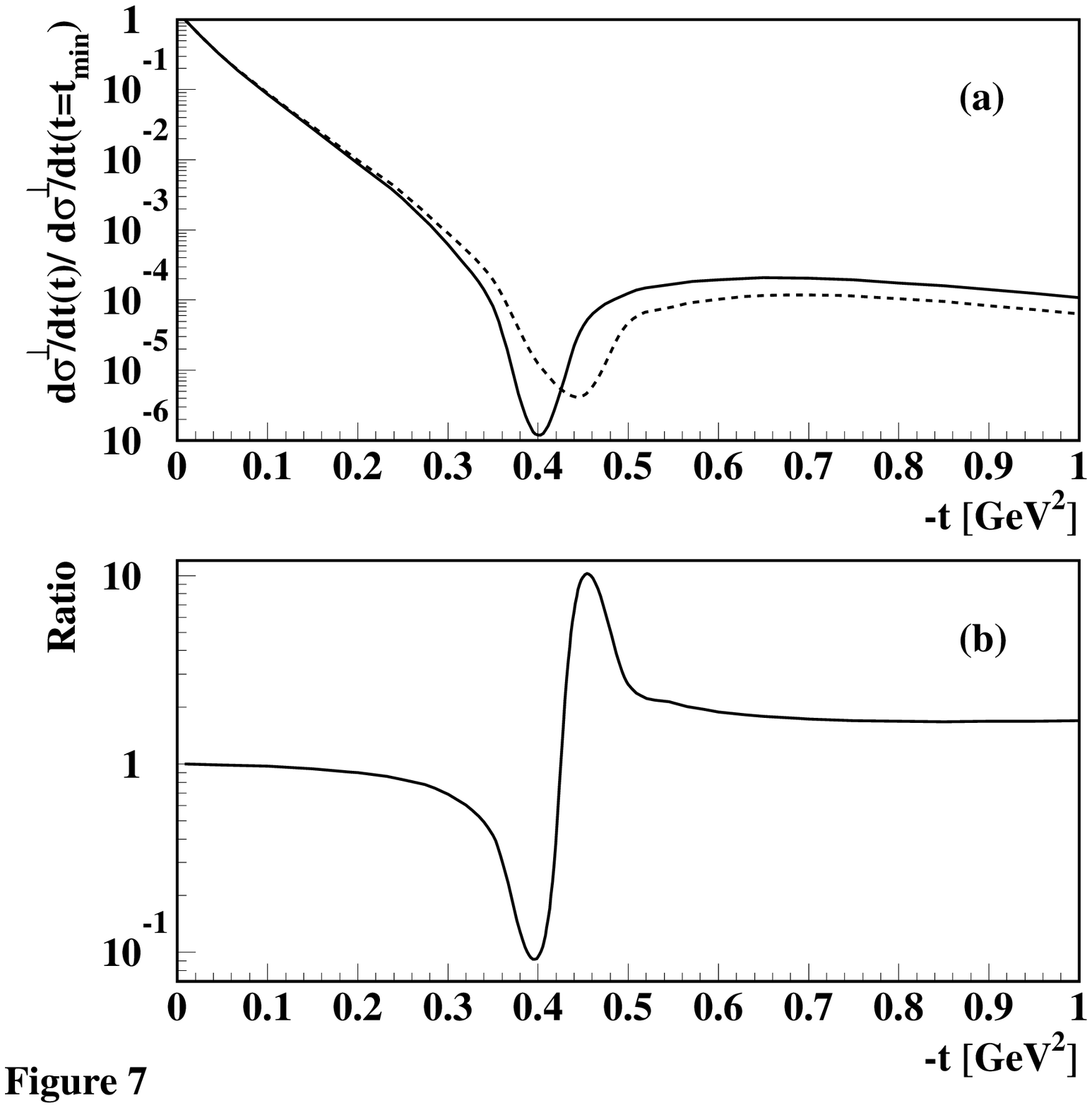}}
\label{Fig.7}
\end{figure}

\end{document}